# Revealing the electronic band structure of trilayer graphene on SiC: An angle-resolved photoemission study

C. Coletti[1,2,*], S. Forti[1], A. Principi[3], K.V. Emtsev[1], A.A. Zakharov[4], K.M. Daniels[5], B.K. Daas[5], M.V.S. Chandrashekhar[5], T. Ouisse[6], D. Chaussende[6], A.H. MacDonald[7], M. Polini[3], U. Starke[1]

[1]*Max-Planck-Institut für Festkörperforschung, Heisenbergstr. 1, D-70569 Stuttgart, DE*

[2]*Center for Nanotechnology Innovation @ NEST, Istituto Italiano di Tecnologia, Piazza San Silvestro 12, I-56127 Pisa, IT*

[3]*NEST, Istituto Nanoscienze-CNR and Scuola Normale Superiore, I-56126 Pisa, IT*

[4]*MAX-lab, Lund University, Lund, S-22100, SE*

[5]*University of South Carolina, 301 S. Main St, Columbia, SC 29208, USA*

[6]*Laboratoire des Matériaux et du Génie Physique-CNRS UMR5628-Grenoble INP, Minatec 3 parvis Louis Néel, 38016 Grenoble, FR*

[7]*Department of Physics, University of Texas at Austin, Austin, Texas 78712, USA*

[*] *camilla.coletti@iit.it*

In recent times, trilayer graphene has attracted wide attention owing to its stacking and electric field dependent electronic properties. However, a direct and well-resolved experimental visualization of its band structure has not yet been reported. In this work, we present angle resolved photoemission spectroscopy (ARPES) data which show with high resolution the electronic band structure of trilayer graphene obtained on α-SiC(0001) and β-SiC(111) via hydrogen intercalation. Electronic bands obtained from tight-binding calculations are fitted to the experimental data to extract the interatomic hopping parameters for Bernal and rhombohedral stacked trilayers. Low energy electron microscopy (LEEM) measurements demonstrate that the trilayer domains extend over areas of tens of square micrometers, suggesting the feasibility of exploiting this material in electronic and photonic devices. Furthermore, our results suggest that on SiC substrates the occurrence of rhombohedral stacked trilayer is significantly higher than in natural bulk graphite.

## I. INTRODUCTION

Recently, a great deal of attention has been devoted to trilayer graphene (TLG) because it displays stacking and electric field dependent electronic properties well-suited for electronic and photonic applications [1-8]. TLG has two naturally stable allotropes characterized by either Bernal (ABA) or rhombohedral (ABC) stacking of the individual carbon layers. In ABA-stacking the atoms of the topmost layer obtain lateral positions exactly above those of the bottom layer (Fig. 1(a)). In an ABC-stacked trilayer each layer is laterally shifted with respect to the layer below by a third of the diagonal of the lattice unit cell (Fig. 1(b)). Several theoretical studies have predicted the electronic dispersion of ABA- and ABC-stacked trilayers using tight-binding approaches [1-3,9-13]. The low-energy band structure of ABA TLG consists of a linearly dispersing (monolayer-like) band and bilayer-like quadratically dispersing bands (Fig. 1(c)) [1,3,11]. Quite differently, ABC trilayers have a single low-energy band with approximately cubic dispersion (Fig. 1(d)) [1-3,12]. A very intriguing distinction between the two allotropes is their behavior in the presence of a perpendicular electric field: ABA-stacked trilayers are expected to display a tunable band overlap, while ABC-stacked trilayers present a tunable band-gap, the latter being very appealing for electronic applications [3,10]. However, the alluring rhombohedral phase is quite rare in nature as the energetically favored Bernal stacking makes up for more than 80% of the existing graphite [14,15].

On the experimental side, progress in revealing the fundamental properties of TLG has been slow as such studies require homogenous trilayers with a well defined stacking sequence over areas of hundreds of micrometers. Infrared conductivity and transport measurements have recently confirmed that a band-gap can be opened in ABC-stacked TLG when applying a perpendicular electric field, while no band-gap has been observed in ABA-stacked trilayers [5]. However, a direct visualization of the electronic band structure of homogenous TLG via angle resolved photoemission spectroscopy (ARPES) has not been reported so far. In 2007, Ohta and colleagues reported ARPES spectra of few-layers graphene on SiC [16]. However, the separation of contributions from areas with different number of layers or different stacking in such a configuration is ambiguous and rather challenging. Clearly, the availability of highly resolved experimental ARPES data for TLG would allow for a direct comparison with the band structure predicted by the tight-binding formalism, thus leading to a precise determination of the interatomic interactions (sketched by the hopping parameters in panel (a)).

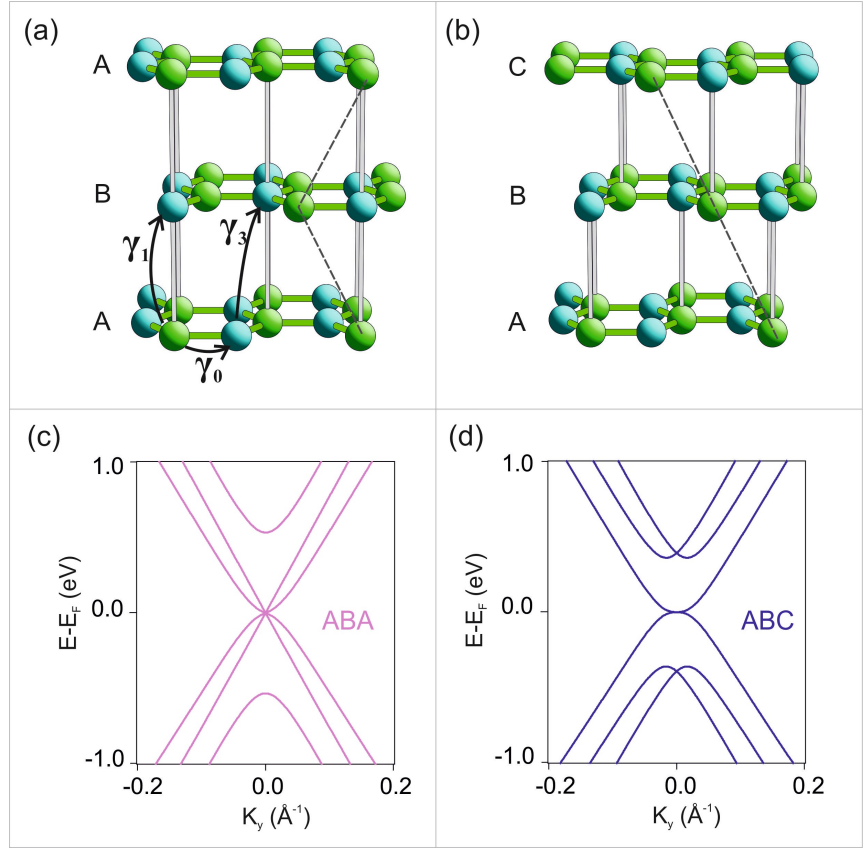

Figure 1. (a,b) Schematic representation of the stacking sequence in (a) Bernal and (b) rhombohedral TLG. The interatomic tight-binding hopping parameters between adjacent layers – thus valid for both stackings – are denoted by the black arrows in panel (a). (c,d) Calculated low-energy band structure for Bernal (c) and rhombohedral (d) TLG.

In the present paper large-area homogenous TLG is obtained on both hexagonal and cubic SiC (i.e., 6H-SiC(0001) and 3C-SiC(111), respectively) by first growing bilayer graphene (BLG) and then adopting the hydrogen intercalation technique described in [17]. The thickness homogeneity of such samples is confirmed by low energy electron microscopy (LEEM) analysis. Having obtained homogenous trilayer graphene we can acquire high-resolution ARPES energy-momentum (E-$k$) maps which correlate well with the band structure calculated by theory for both ABA and ABC stacks. We use band structure results obtained from tight-binding calculations to fit the experimental ARPES data and to extract the hopping parameters both for ABA- and ABC-stacked trilayers. Remarkably, the analysis of the ARPES data suggests that, on both 6H-SiC(0001) and 3C-SiC(111) substrates, graphene exhibits a tendency towards the development of ABC type stacking that is noticeably higher than that observed in natural graphite. Growing trilayer graphene on SiC substrates might therefore be the answer to the challenge of controllably synthesizing ABC-stacked trilayers [8].

## II. METHODS

In our experiments, homogeneous graphene bilayers were grown on nominally on-axis oriented 6H-SiC(0001) substrates and on highly homogenous free-standing 3C-SiC(111) epilayers [18]. The growth parameters were finely optimized to obtain the highest bilayer coverage. Growth on the hexagonal polytype was performed in an inductively heated RF furnace at a temperature of 1350 °C, a pressure of $10^{-5}$ mbar for 1 hour [19]. On the cubic polytpe, thermal annealing was performed at a temperature of 1600 °C, an Ar pressure of 100 mbar, for 20 minutes. H-intercalation was performed by annealing the samples for 20 to 40 minutes in a hydrogen atmosphere at a pressure of 830 mbar and a temperature of 1000 °C. The thickness and homogeneity of the as-grown and hydrogen intercalated samples was evaluated via LEEM using the ELMITEC3 instrument at the end-station of beamline I311 at MAX-LAB. The electronic dispersion was investigated via ARPES at the end-station of the SIS beamline at the Swiss Light Source synchrotron facility using p-polarized light. The spectra and the costant energy maps (CEMs) reported were measured with a photon energy of 90 eV.

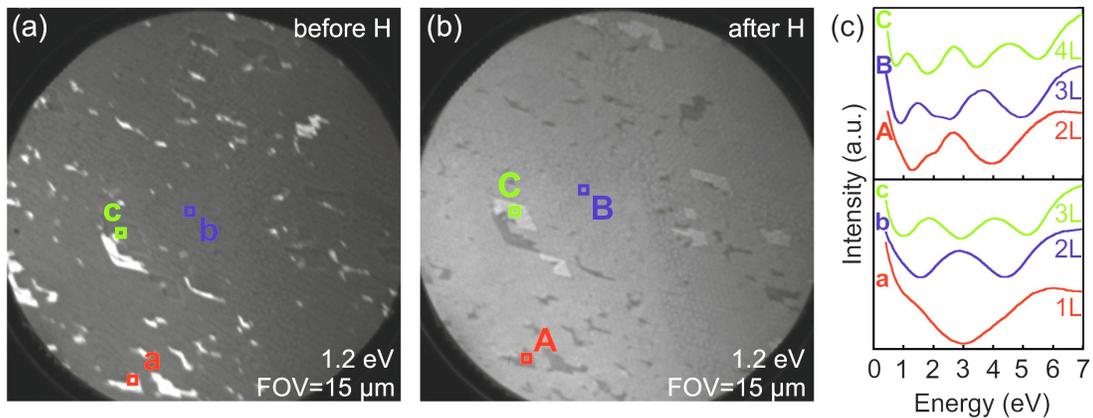

Figure 2. Representative LEEM micrographs with a field of view of 15 μm recorded with an electron energy of 1.2 eV for (a) as-grown BLG on 6H-SiC(0001) and (b) the same area after hydrogen intercalation. (c) Electron reflectivity curves collected for the labeled regions (in panels a and b) of the initial surface (bottom graph) and of the hydrogen intercalated graphene sample (top graph).

## III. RESULTS

A characteristic LEEM micrograph for as-grown BLG on 6H-SiC(0001) is displayed in Figure 2(a) with a field of view (FOV) of 15 μm. At the energy of 1.2 eV used for recording the image, regions with different graphene thickness can be distinguished by differences in the reflected intensity. Although surface domains with three different grayscale contrasts can be identified, the sample is highly homogeneous with the medium gray domains (label b) occupying more than 80% of the overall area. The number of dips in the electron reflectivity spectra plotted in the bottom of panel (c) confirms that these areas consist of BLG while the small regions with light-gray (label a) and dark-gray (label c) contrast are monolayer graphene (MLG) and TLG, respectively [20]. The band structure of the sample was measured around the –point of the graphene Brillouin zone (BZ) using synchrotron radiation based ARPES. The spectrum shown in Fig. 3(a) is representative of the entire area of the sample. The spectrum is extremely sharp and exclusively consists of parabolic bands, the signature of bilayer graphene, corroborating the extreme homogeneity of the graphene film. Hence, the graphene thickness is essentially constant over a large area (the spot-size of the UV light beam is about 100x50 μm$^2$) and the small percentage of domains of different thickness observed via LEEM does not cause significant contributions to the measured band structure. In Fig. 3(d), theoretical bands obtained by tight-binding calculations for a Bernal stacked bilayer using the formalism of McCann and Fal'ko [21] are fitted to the experimental data (see Supplemental Material [22]). As expected for epitaxial BLG on SiC, the Fermi level is shifted by around 0.3 eV above the Dirac energy of the π-bands – indicative of n-type doping [23,24]. Also, the characteristic band-gap of  120 meV caused by the electrostatic asymmetry of the bilayer slab on the SiC substrate is visible [23,24]. The LEEM micrograph in Fig. 2(b) shows the same sample area as in Fig. 2(a), yet, upon annealing the sample in hydrogen. As described in [17,25], this treatment causes hydrogen to intercalate between the buffer layer and the SiC(0001) surface. Hydrogen atoms passivate the Si dangling bonds, so that the overlaying graphene layers are electronically and structurally decoupled from the SiC substrate. In this way, the buffer layer becomes an electronically active monolayer and, more generally, a n-layer graphene film transforms into a (n+1) layer graphene film. Indeed, the number of dips in the electron reflectivity spectra reported in the upper panel of Fig.2(c) confirms the conversion of all the n-layers into (n+1)-layers. X-ray photoelectron spectroscopy (XPS) analysis also confirms the complete intercalation (Supplemental Material [22]). Thus, after hydrogen intercalation, the sample consists of highly homogenous quasi-free standing trilayer graphene (QFTLG).

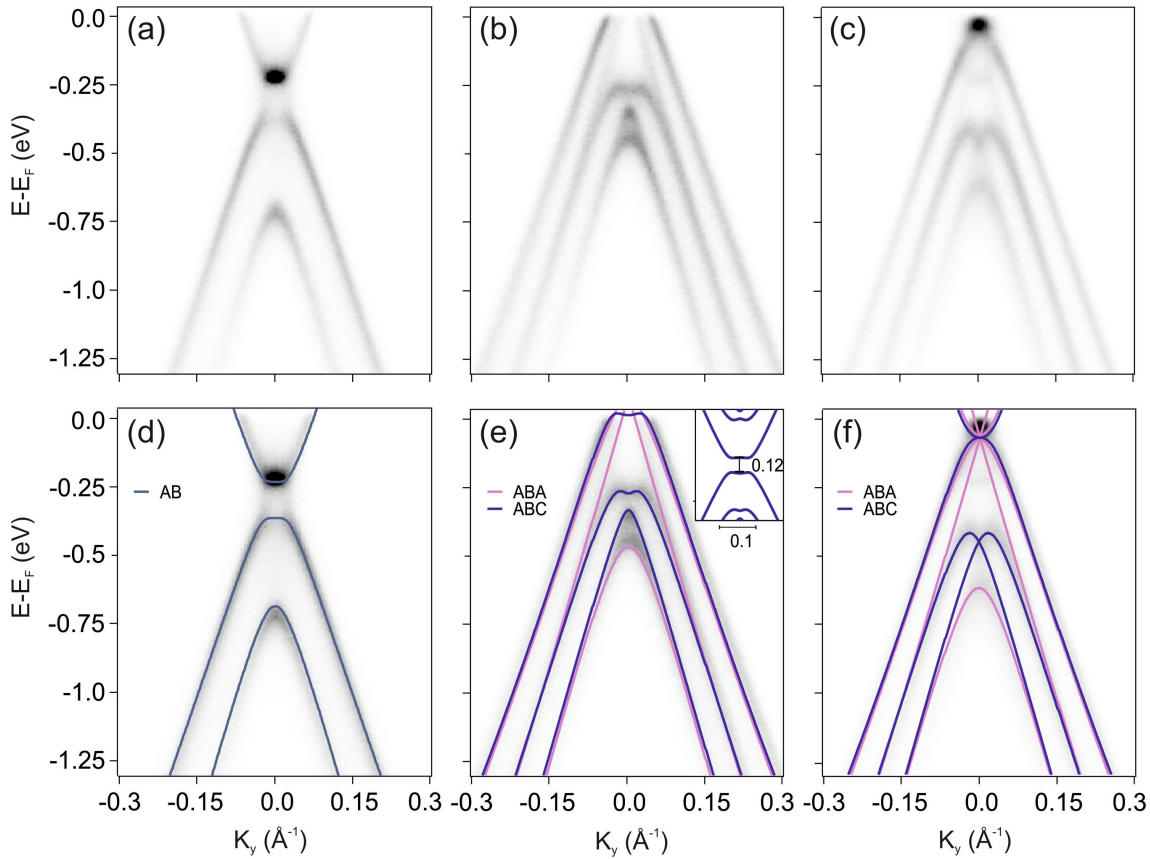

Figure 3. Dispersion of the π-bands measured via ARPES for as-grown BLG on 6H-SiC(0001) (a), QFTLG annealed at 400 °C (b) and at about 800° C (c). The spectra are measured with a photon energy of 90 eV and with scans oriented perpendicular to the -direction of the graphene Brillouin zone. (d-f) Tight-binding bands fitted to the experimental data shown in (a), (b) and (c), respectively. The fitting retrieves a band-gap in the ABC dispersion in panel (b) of 120 meV (inset in panel (e)).

From this sample we have acquired the first well-resolved direct visualization of the electronic band structure of TLG as displayed in figure 3(b). The spectrum shown was collected after outgassing the sample at 400 °C, a temperature sufficient to remove air contamination but well below the onset of hydrogen desorption [25]. A mixture of several sharp bands can be observed. The high quality of the measured band structure allows for a precise identification of the trilayer stacking sequence. To this end, theoretical bands derived from tight-binding Hamiltonians describing the ABA and ABC trilayers were fit to the experimental data (see Supplemental Material [22]). Panel (e) shows the results of the fitting procedure superimposed to the electronic dispersions obtained experimentally. The two stacking sequences, ABA and ABC, can be clearly distinguished as indicated by the respective pink and blue colored fitting curves. The accurate overlap of the calculated bands with the experimental data reveals unambiguously that QFTLG on SiC contains domains of both Bernal and rhombohedral stacking, in contrast to natural graphite which typically only features ABA stacking [14,15]. The excellent fit also indicates that all experimentally visible bands belong to trilayer graphene, thus corroborating the overall homogeneous graphene thickness. From the fits in panel (e) the Dirac energy can be determined to be about 90 meV above the Fermi energy. The p-type doping is typical for hydrogen intercalated samples on α-SiC [26] and has been recently attributed to the spontaneous polarization of the substrate imposed by hexagonal SiC's pyroelectricity [27]. This polarization obviously induces an electrostatic field across the trilayer slab (the on-site Coulomb potential difference between the first and the third layer is calculated to be 0.12 eV) which modifies the band structure of trilayer graphene as described in [3,10]. In particular, from the fits it can be derived that at the -point an energy band-gap of 120 meV is induced (inset in panel (e)). This value indeed is in agreement with results from infrared conductivity measurements [5]. The error bar for the band-gap is estimated to be ± 25 meV (Supplemental Material [22]). As reported in Refs. [17,25], by annealing a quasi-free standing monolayer graphene (QFMLG) sample at higher temperatures it is possible to achieve charge neutrality within a few meV. This is also successful for the present QFTLG sample. The band structure shown in panel (c) was measured after prolonged annealing at about 800 °C, which is a higher temperature than that needed to obtain charge neutral quasi-free mono- and bilayer graphene [28]. In fact, the sample appears to have acquired a minimal n-doping after this treatment by possibly desorbing an excess of hydrogen from the Si dangling bonds. The visibility of the onset of the conduction band allows one to appreciate the absence of a measurable band-gap. Hence, after annealing and in consequence retrieving charge neutrality, no on-site Coloumb potential difference is necessary for the calculated bands to be superimposed onto the experimental data in panel (f).

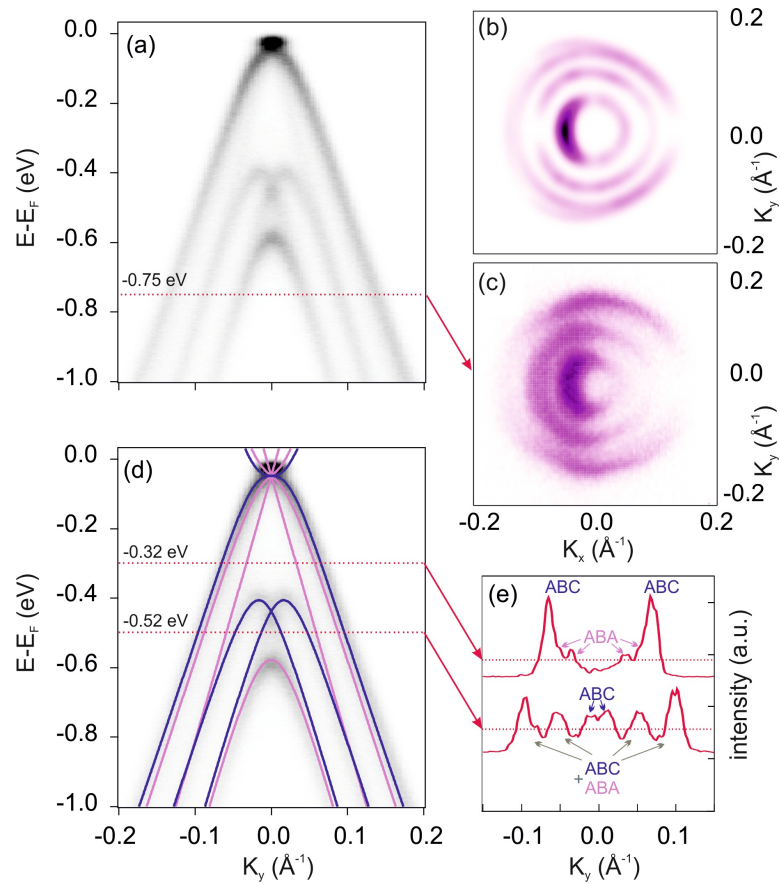

Figure 4. Electron dispersion spectra measured via ARPES for QFTLG on 3C-SiC(111): raw data (a) and superimposed tight-binding bands (d). Theoretical (b) and experimental (c) constant energy maps at -0.75 eV. (e) MDCs measured at energies of -0.32 eV and -0.52 eV. The spectra and the CEM are measured with a photon energy of 90 eV.

High-quality QFTLG could also be obtained on 3C-SiC(111) substrates as demonstrated by the ARPES spectra in Figure 4(a). This is a remarkable accomplishment, considering that until recently even the growth of large area MLG was considered to be a challenge [29]. Panel (a) demonstrates the sharpness of the bands and the absence of contributions from domains of different thicknesses, while panel (d) shows the fitted bands superimposed to the raw data . Similar to the case of QFMLG reported in [29], on the cubic substrate graphene is almost charge neutral without the need for annealing. This finding once more confirms the doping model suggested in [27], as spontaneous polarization does not occur in 3C-SiC(111) substrates due to the cubic symmetry. A small n-type doping of about $7 \times 10^{11}$ cm$^{-2}$ can be derived from the onset of the conduction band being about 40 meV below the Fermi level. This observation can be explained by residual defects present at the SiC/graphene interface inducing local negative image charge from the Si dangling bonds.

Notably, a simple visual inspection suggests that the intensities of the ABC bands of all the measured spectra are higher than those of the ABA contributions. This is illustrated by representative momentum distribution curves (MDCs) plotted in Fig. 4(e). The MDCs were measured at -0.32 and -0.52 eV, energies at which the contributions of the two stackings can be separately distinguished. Of course, it must be taken into consideration that the photoemission intensity of single ABA and ABC branches is expected to vary as a consequence of varying strength and direction of interatomic interactions [16,30]. Nevertheless, the MDCs clearly indicate that the ABC branches are significantly more intense than the ABA ones. Also, we note that by measuring at photon energies different from 90 eV, we obtained spectra with varying ABA and ABC contributions. Yet, the intensities of the rhombohedral bands were never smaller than those of the Bernal stacking. These results suggest that ABC type of stacking occurs in QFTLG on SiC with a significantly higher incidence than in nature. The tendency of graphene to form on SiC in ABC-stacking could be explained by a weakening of the $\gamma_5$ interatomic interaction – a major contributor to the stability of the ABA stacking – due to the displacement of carbon atoms in the buffer layer during the growth process [31].

| Substrate | Stacking | γ | γ | γ |
|---|---|---|---|---|
| 6H-SiC(0001) | ABC | -2.86 | -0.38 | -0.24 |
| 6H-SiC(0001) | ABA | -3.05 | -0.39 | -0.20 |
| 3C-SiC(111) | ABC | -3.24 | -0.39 | -0.24 |
| 3C-SiC(111) | ABA | -3.5 | -0.37 | -0.20 |

Table I. Hopping parameters for Bernal and rhombohedral stacked TLG on hexagonal and cubic SiC substrates directly calculated from tight-binding fits to experimental ARPES data. All values are in eV.

The hopping parameters obtained from fitting the experimental spectra for both crystallographic arrangements of trilayer graphene on 6H-SiC(0001) and 3C-SiC(111) are listed in Table I. In our calculation we considered the nearest neighbor intralayer and interlayer coupling terms $\gamma_0$ and $\gamma_1$, and the next nearest-neighbor interlayer coupling term $\gamma_3$. Weaker coupling terms and tunneling processes describing next-neighbor hopping were neglected since their change has no noticeable effects on the resulting bands. The absolute values obtained for $\gamma_0$ and $\gamma_1$ agree well with those predicted by theory [2,10,11,13] and experimentally retrieved for few layer graphene and graphite [5,16,32]. The anisotropy of photoemission constant-energy maps (CEMs) measured experimentally (see Figure 4(c)) depends on the magnitude and relative sign of the interlayer coupling parameters ($\gamma_1$ and $\gamma_3$) [30]. Thus, by comparing experimental and theoretical (see Supplemental Material [22]) CEMs it is possible to extract the sign of the interlayer coupling parameter $\gamma_1$ and the relative sign between $\gamma_1$ and $\gamma_3$. By adopting this procedure we found that $\gamma_1 < 0$ and $\gamma_3 < 0$ for both stacking arrangements. Indeed, the agreement between the calculated and the experimental CEMs, as exemplified in Figures 4(b) and (c), is striking. It should be noted that, although the sign of $\gamma_1$ has a directly observable effect on the ARPES bands, up to now it has often been assumed to be positive [2,5,10,11,13,16,32]. As suggested in [30], the negative sign should be a natural consequence of the z→-z asymmetry of the $p_z$ orbitals of carbon. The term $\gamma_3$, which defines the strength of the trigonal warping effect, is in agreement with what is predicted by theory and experimentally obtained for graphite [2,13,32].

The values of the hopping parameters are quite similar for our cubic and hexagonal substrates with the exception of $\gamma_0$, which is higher for graphene on 3C-SiC(111). From $\gamma_0$ we can derive that the band velocity of the rhombohedral QFTLG on 6H-SiC(0001) is about $0.93 \times 10^6$ m/s, while on 3C-SiC(111) it is calculated to be about $1.05 \times 10^6$ m/s. We note that a distinctively high band velocity was found also for QFMLG on 3C-SiC(111) [29], suggesting a dependence of the Fermi velocity from the substrate as already reported in [33]. However, control experiments indicate that the differences in band velocity rather arise from a different concentration of scattering centers due to surface morphology (Supplemental Material [22]).

## IV. CONCLUSION

In summary, we have demonstrated that high quality QFTLG can be obtained on both cubic and hexagonal SiC substrates. We have directly visualized – via ARPES – extremely sharp electron dispersion spectra of ABA and ABC stacked trilayers and shown that they correlate well with the tight-binding calculations reported so far. For ABC stacks and in the presence of an electrostatic asymmetry, we detect the existence of a band-gap of 120 meV, which makes this graphene structure appealing for electronic applications. Using a tight-binding approach, we provide a direct determination of the relevant hopping parameters. Furthermore, we observe that QFTLG on SiC presents an occurrence of the ABC type of stacking with a higher percentage than observed in natural graphite. Hence, TLG on SiC might be the material of choice for the fabrication of a new class of gap-tunable devices.

**ACKNOWLEDGMENTS**

C.C. acknowledges the Alexander von Humboldt Foundation for financial support. This work was supported by the Deutsche Forschungsgemeinschaft in the framework of the Priority Program 1459 Graphene (Sta315/8-1). This research was partially funded by the European Community's Seventh Framework Programme: Research Infrastructures (FP7/2007-2013) under grant agreement no. 226716. Support by the staff at MAX-Lab (Lund, Sweden) and SLS (Villigen, Switzerland) is gratefully acknowledged.

References

[1] H. Min, and A. H. MacDonald, *Prog. Theor. Phys. Suppl*. **176**, 227 (2008).
[2] F. Zhang, B. Sahu, H. Min and A. H. MacDonald, *Phys. Rev. B* **82**, 035409 (2010).
[3] M. Koshino, *Phys. Rev. B* **81**, 125304 (2010).
[4] M. F. Craciun, S. Russo, M. Yamamoto, J. B. Oostinga[2,3], A. F. Morpurgo, and S. Tarucha, *Nature Nanotechn*. **4**, 383 - 388 (2009).
[5] C. H. Lui, Z. Li, K. F. Mak, E. Cappelluti, and T. F. Heinz, *Nature Phys*. **7**, 944–947 (2011).
[6] W. Bao, L. Jing, J. Velasco Jr, Y. Lee, G. Liu, D. Tran, B. Standley, M. Aykol, S. B. Cronin, D. Smirnov, M. Koshino, E. McCann, M. Bockrath, and C. N. Lau, *Nature Phys*. **7**, 948–952 (2011).
[7] L. Zhang, Y. Zhang, J. Camacho, M. Khodas, and I. Zaliznyak, *Nature Phys*. **7**, 953–957 (2011).
[8] A. Yacoby, *Nature Phys*. **7**, 925-926 (2011).
[9] F. Guinea, A. H. Castro Neto, and N. M. R. Peres, *Phys. Rev. B* **73**, 245426 (2006).
[10] M. Aoki, and H. Amawashi, *Solid State Communications* **142,** 123–127 (2007).
[11] A. Grüneis, C. Attaccalite, L. Wirtz, H. Shiozawa, R. Saito, T. Pichler, and A. Rubio, Phys. Rev. B **78**, 205425 (2008).
[12] M. Koshino, and E. McCann, *Phys. Rev. B* **80**, 165409 (2009).
[13] A. A. Avetisyan, B. Partoens, and F. M. Peeters, *Phys. Rev. B* **81**, 115432 (2010).
[14] H. Lipson, and A. R. Stokes, *Proc. R. Soc. A* **101**, *181* (1942).
[15] C. H. Lui, Z. Li, Z. Chen, P. V. Klimov, L. E. Brus, and T. F. Heinz, *Nano Lett*. *11*, 164–169 (2011).
[16].T. Ohta, A. Bostwick, J. L. McChesney, T. Seyller, K. Horn, and E. Rotenberg, *Phys. Rev. Lett*. **98**, 206802 (2007).
[17] C. Riedl, C. Coletti, T. Iwasaki, A. A. Zakharov, U. Starke, *Phys. Rev. Lett*. **103**, 246804 (2009).
[18] D. Chaussende, L. Latu-Romain, L. Auvray, M. Ucar, M. Pons, and R. Madar, *Mater. Sci. Forum* **483-485**, 225 (2005).

[19] B. K. Daas, K. M. Daniels, T. S. Sudarshan, and M. V. S. Chandrashekhar, *J. Appl. Phys.* **110**, 113114, (2011).

[20] H. Hibino, H. Kageshima, F. Maeda, M. Nagase, Y. Kobayashi, and H. Yamaguchi, *Phys. Rev. B* **77**, 075413 (2008).

[21] E. McCann, and V. I. Fal'ko, *Phys. Rev. Lett.* **96**, 086805 (2006).

[22] See Supplemental Material at URL for details about: carbon core level analysis, tight-binding parameters, fitting procedures, constant energy maps, band velocity dependence.

[23] T. Ohta, A. Bostwick, T. Seyller, K. Horn, and E. Rotenberg, *Science* **313**, 951 (2006).

[24] C. Coletti, C. Riedl, D. S. Lee, B. Krauss, K. von Klitzing, J. Smet, and U. Starke, *Phys. Rev. B* **81**, 235401 (2010).

[25] S. Forti, K. V. Emtsev, C. Coletti, A. A. Zakharov, and U. Starke, *Phys. Rev. B* **84**, 125449 (2011).

[26] C. Coletti, S. Forti, K. V. Emtsev, and U. Starke, *GraphITA 2011*, pp. 39-49, Springer Berlin Heidelberg (2012).

[27] J. Ristein, S. Mammadov, and T. Seyller, *Phys. Rev. Lett.* **108**, 246104 (2012).

[28] C. Riedl, C. Coletti, and U. Starke, *J. Phys. D: Appl. Phys.* **43** 374009 (2010).

[29] C. Coletti, K. V. Emtsev, A. A. Zakharov, T. Ouisse, D. Chaussende, and U. Starke, *Appl. Phys. Lett.* **99**, 081904 (2011).

[30] M. Mucha-Kruczyński, O. Tsyplyatyev, A. Grishin, E. McCann, V. I. Fal'ko, A. Bostwick, and E. Rotenberg, *Phys. Rev. B* **77**, 195403 (2008).

[31] W. Norimatsu, and M. Kusunoki, *Phys. Rev. B* **81**, 161410 (2010).

[32] L. M. Malard, J. Nilsson, D. C. Elias, J. C. Brant, F. Plentz, E. S. Alves, A. H. Castro Neto, and M. A. Pimenta, *Phys. Rev. B* 76, 201401(R) (2007).

[33] C. Hwang, D. A. Siegel, S.-K. Mo, W. Regan, A. Ismach, Y. Zhang, A. Zettl, and A. Lanzara, *Scien. Rep.* **2**, 509 (2012).